%% file: 00_main.tex
\documentclass[journal, preprint]{vgtc}                     % final (journal style)

\pdfoutput=1 
\onlineid{0}

\ieeedoi{xx.xxxx/TVCG.201x.xxxxxxx}

\vgtccategory{Research}

\title{Are We Closing the Loop Yet? Gaps in the Generalizability of VIS4ML Research}

\author{%
 \authororcid{Hariharan Subramonyam}{0000-0002-3450-0447}, and 
 \authororcid{Jessica Hullman}{0000-0001-6826-3550}
}

\authorfooter{
  \item
  	Hariharan Subramonyam is with Stanford University.
  	E-mail: harihars@stanford.edu.
  \item
  	Jessica Hullman is with Northwestern University.
  	E-mail: jhullman@northwestern.edu.
}

\abstract{ Visualization for machine learning (VIS4ML) research aims to help experts apply their prior knowledge to develop, understand, and improve the performance of machine learning models. In conceiving VIS4ML systems, researchers characterize the nature of human knowledge to support human-in-the-loop tasks, design interactive visualizations to make ML components interpretable and elicit knowledge, and evaluate the effectiveness of human-model interchange. We survey recent VIS4ML papers to assess the generalizability of research contributions and claims in enabling human-in-the-loop ML. Our results show potential gaps between the current scope of VIS4ML research and aspirations for its use in practice. We find that while papers motivate that VIS4ML systems are applicable beyond the specific conditions studied, conclusions are often overfitted to non-representative scenarios, are based on interactions with a small set of ML experts and well-understood datasets, fail to acknowledge crucial dependencies, and hinge on decisions that lack justification. We discuss approaches to close the gap between aspirations and research claims and suggest documentation practices to report generality constraints that better acknowledge the exploratory nature of VIS4ML research. 
}

\keywords{VIS4ML, Visualization, Machine learning, Human-in-the-loop, Human Knowledge, Generalizability, Survey. }

\graphicspath{{figs/}{figures/}{pictures/}{images/}{./}} % where to search for the images

\usepackage{tabu}                      % only used for the table example
\usepackage{booktabs}                  % only used for the table example
\usepackage{lipsum}                    % used to generate placeholder text
\usepackage{mwe}                       % used to generate placeholder figures
\usepackage{balance}
\usepackage{mathptmx}                  % use matching math font
\usepackage{comment}

\usepackage{tabularray}
\UseTblrLibrary{booktabs}
\usepackage{stfloats}
\usepackage{amsmath}%
\usepackage{MnSymbol}%
\usepackage{wasysym}%
\newcommand*{\greysquare}{\textcolor[HTML]{488A99}{$\blacksquare$}}

\newcommand*{\HITLsquare}{\textcolor[HTML]{BA9044}{$\blacksquare$}}

\newcommand*{\vissquare}{\textcolor[HTML]{6AB187}{$\blacksquare$}}

\usepackage{makecell}

\begin{document}

%%%%%%%%%%%%%%%%%%%%%%%%%%%%%%%%%%%%%%%%%%%%%%%%%%%%%%%%%%%%%%%%
%%%%%%%%%%%%%%%%%%%%%% START OF THE PAPER %%%%%%%%%%%%%%%%%%%%%%
%%%%%%%%%%%%%%%%%%%%%%%%%%%%%%%%%%%%%%%%%%%%%%%%%%%%%%%%%%%%%%%%


\maketitle

\input{01_intro}
\input{02_relatedwork}

\input{03_methods}

\input{04_results}

\input{06_discussion}
\input{07_conclusion}

\bibliographystyle{abbrv-doi}
\bibliography{99_refs}

\end{document}

%% file: 01_intro.tex
\section{Introduction}

Visualization for machine learning (VIS4ML) research aims to support human involvement in the machine learning (ML) process by making ML models interpretable to humans~\cite{sacha2018VIS4ML}. The underlying assumption is that by providing experts such as ML engineers and domain specialists with appropriate \textit{visual representations} of the modeling pipeline, they will be able to combine their relevant \textit{prior knowledge} with the machine representation toward positive ends - i.e., human-in-the-loop (HITL) machine learning. For instance, DataDebugger~\cite{xiang2019interactive} supports human correction of mislabeled training data, INFUSE~\cite{krause2014infuse} enables domain expert involvement in feature engineering, and ConceptExplainer~\cite{huang2022conceptexplainer} allows analysts to extract concept-based explanations for explainable AI tasks. 
In an ideal scenario, the knowledge generated through bespoke VIS4ML contributions can influence real-world ML workflows, in which practitioners can use these tools to interpret and develop performant models. 

In this paper, we consider evidence of the \textit{generalizability} of VIS4ML research. Generalizability concerns the alignment between the general claims made about the effectiveness and applicability of VIS4ML contributions and the quantitative or qualitative evidence presented to validate those claims. 
In VIS4ML research, this depends on how \textit{design hypotheses} -- propositions about effective visualization and interaction choices for meeting HITL task needs -- are operationalized in light of the researchers' generalization goals.
This includes understanding how papers go from aspirations about how VIS4ML systems can enable HITL tasks, to specific intended effects of involving human knowledge in a pipeline, to particular design instantiations meant to realize these effects, to the evaluation of those design artifacts, to the conclusions that are ultimately drawn about effective VIS4ML strategies. When design hypotheses involve unstated assumptions that are overlooked in interpreting the results--for example, about the degree of knowledge people have of model components--we should not expect claims to generalize to settings where those assumptions are not in place. Further, we would expect claims entailed by the design hypotheses to be directly validated by evidence of use, including evaluating the validity and robustness of human-generated insights and how these ultimately affect the target learning pipeline or downstream outcomes. However, causal inferences that users generate to explain model performance may not be verifiable without further data collection, or researchers may lack visibility into the larger lifecycle of a model that they intend to affect. When research claims are based on unstated dependencies and overlooked gaps in evidence, the claimed effects of applying VIS4ML contributions are unlikely to realize in practice.

In this work, we critically examine a set of 52 VIS4ML papers to characterize the space of design hypotheses and evaluation practices and identify gaps that could hinder the adoption of VIS4ML research in practice. Our analysis surfaces patterns in how papers envision the role of human knowledge in VIS4ML, the knowledge assumptions made of system users, the algorithms and interpretability approaches they rely on, and the approaches they take to evaluate their hypotheses. We find that a majority of VIS4ML papers aim to combine visualizations and interactivity to bring about concrete improvements to an ML pipeline. Yet, more often than not, these improvements are not directly evaluated. We also identify common dependencies, for example, on the same small group of experts during development and evaluation, on well-known datasets, and on post-hoc interpretability methods that often lack faithfulness guarantees --- that may threaten the ability of independent authors or practitioners to experience the same gains when they apply the contributed approaches in related contexts.

Broadly, our analysis finds the current scope of human-in-the-loop VIS4ML is somewhat limited and draws attention to potential threats to the practical adoption of VIS4ML contributions and the generalizability of research claims. To bridge the gap between bespoke VIS4ML contributions and its use in ML production workflows, we make short and longer-term recommendations for action, including transparent documentation of unstated assumptions and constraints, tightening loose derivation chains in the logical progression from aspirations for human knowledge integration to designs and their evaluation, and exploring partnerships with the broader human-centered AI research community.  

%% file: 02_relatedwork.tex
\section{Background}

\subsection{Taxonomizing VIS4ML} \label{sec:taxonomy}

Existing surveys of visual analytics for ML research~\cite{endert2017state,hohman2018visual,sacha2016human,sacha2018VIS4ML,sperrle2020should,sperrle2021survey,yuan2021survey} taxonomizes goals, activities, and human inputs to the modeling pipeline (separate from the indirect use of ML for improving visual analytics pipelines, e.g.,~\cite{brown2016human}). This includes data quality and feature engineering before model building, understanding and diagnosing issues with parameters or training dynamics during model fitting and selection, and reasoning about results after model building. Taxonomies also capture differences in intended audiences for VIS4ML tools, from ML experts to non-experts or domain experts~\cite{hohman2018visual,sacha2016human,yuan2021survey}, and commonly used visualization and interaction techniques ~\cite{hohman2018visual,sacha2016human}.

Most relevant to our work is Sperlle et al.'s~\cite{sperrle2021survey} survey of human-centered evaluations of human-centered machine learning, which characterizes heterogeneity in evaluation styles and assesses data types, analysis tasks, and interactivity in VIS4ML. They differentiate the knowledge requirements of VIS4ML users, including ML versus domain expertise.   
However, because their scope is broader than ours (nearly half of the 71 papers they survey study explainable AI techniques in lab settings, similar to several other recent surveys~\cite{mohseni2021multidisciplinary,sperrle2020should}), their results are less targeted to visualization-specific research.
Additionally, the aim of our work is unique 
in that, we are interested in the alignment between researchers' aspirations and claims and their methods, including how knowledge is claimed to be produced through moving from research aspirations to specific design hypotheses to the validation of those hypotheses. Hence we focus on the epistemic status of VIS4ML and the generalizability of results than prior surveys.

\subsection{Knowledge generation through visual data analysis}
We investigate the forms of knowledge and insight authors describe to motivate and evaluate VIS4ML systems. Our work relates to knowledge generation (KG) models used in visual analytics~\cite{endert2014human,federico2017role,sacha2014knowledge}, which aim to explain the process by which analysts generate knowledge in working with interactive visualizations of data or models. For instance, Sacha et al.'s KG model~\cite{sacha2014knowledge} adapts sensemaking concepts to describe how an analyst engages in iterative exploration and verification loops with an interactive visualization system to generate knowledge. The process is conceptualized through loops in which the analyst takes \textit{actions}, referring to tasks that generate tangible, unique responses from the visual analytics system, to explore visualized evidence for \textit{findings}, or visual patterns, perhaps driven by an analytical goal. The identification of \textit{findings} leads either to further interaction with the system or to new \textit{insights} when the analyst applies their prior knowledge to interpret the results within the domain-specific setting. 

Specific to VIS4ML, Sacha et al.~\cite{sacha2018VIS4ML} contribute an ontology that breaks complex sequences of human interactions with a VIS4ML system into K-Driven processes, which take in human input to control the process, K-Oriented processes, which output information for humans to process, and K-Centered processes, which are designed for human interaction and cannot be easily further broken down. While they contribute a language for representing interactions, we empirically investigate how such processes are studied in the VIS4ML literature and the sorts of human knowledge and capabilities they assume. Others have studied the variety of techniques by which prior knowledge can be integrated into machine learning systems~\cite{hong2020human,kerrigan2021survey,von2019informed}, for example, how the integration of domain knowledge in machine learning pipelines more broadly is often informal and under-described in applied ML research~\cite{kerrigan2021survey}, which our results corroborate for VIS4ML.

\subsection{Challenges in Evaluating Human-in-the-loop ML}

Prior work has noted challenges in choosing success metrics for VIS4ML tools to identify whether a human-machine collaboration is successful~\cite{boukhelifa2020challenges}. When metrics or ``signals'' of performance that a human-in-the-loop system surfaces are locally relevant but poorly connected to the downstream application for which the model is intended, then human attempts to optimize performance for these metrics, such as by cleaning input data, may not effect or even hurt downstream outcomes~\cite{neutatz2021cleaning}. Similarly, when the specific contributions of a human versus an automated component are not well defined, it is difficult to identify the proper evaluation for the research claims~\cite{boukhelifa2020challenges,sperrle2021survey}.

Our aim to uncover hidden dependencies in VIS4ML aligns with calls for more reproducible, replicable, and robust ML research~\cite{haibe2020transparency}, especially ML system evaluations~\cite{liao2021we}.  Similar to the replication crisis in experimental research, overlooked dependencies, sources of variance, and conventions that encourage bold claims can lead researchers to misattribute performance differences to parts of an ML pipeline~\cite{hullman2022worst}.

%% file: 03_methods.tex
\section{Methodology}

%%%%%%%%%%%%%%%%%%%%%%%%%%%%%%%%%%%%%%%%%%%%%%%%%%%%%%
\begin{figure*}[ht!]
\centering
\includegraphics[width=\textwidth]{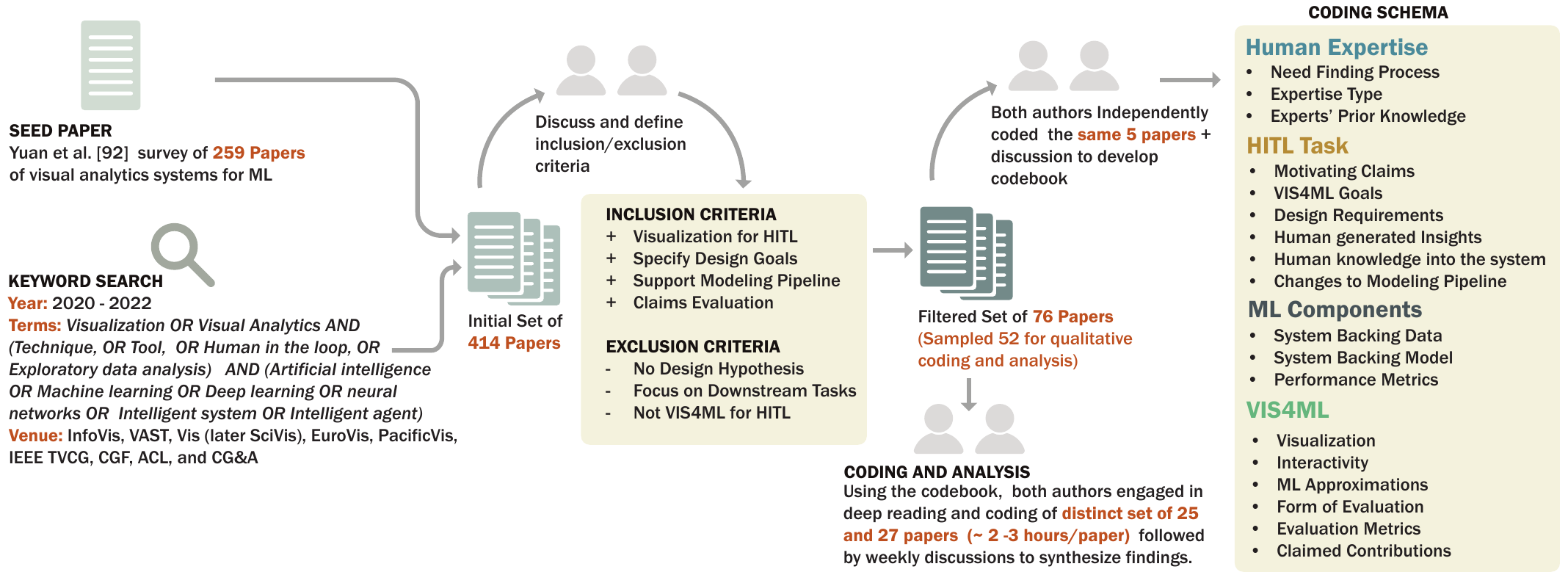}
\caption{Method for Paper Selection, Codebook Development, and Coding and Analysis. }
\label{figure:coding}
\end{figure*}
%%%%%%%%%%%%%%%%%%%%%%%%%%%%%%%%%%%%%%%%%%%%%%%%%%%%%%
To characterize the nature of VIS4ML research contributions with a particular emphasis on human-in-the-loop machine learning, we conducted a qualitative survey of recent research papers proposing and evaluating VIS4ML tools. Figure~\ref{figure:coding} shows an overview of our six month-long paper collection process, analysis, and discussion of findings. 

\subsection{Paper Selection}
We took a two-fold approach to identify papers of interest. First, we seeded our list of papers with Yuan et al.'s set of 259 papers published between 2010-2020 at InfoVis, VAST, Vis (later SciVis), EuroVis, PacificVis, IEEE TVCG, CGF, and CG\&A~\cite{yuan2021survey} used in their survey of \textit{VA techniques} for ML. Second, we applied a keyword search to retrieve papers between 2020-2022 from the same venues. Our keyword selection was expansive and included visualization-specific terms such as \textit{visualization, visual analytics, human-in-the-loop, exploratory data analysis, techniques, tools}, and ML keywords, including \textit{artificial intelligence, machine learning, neural network, deep learning, intelligent system, and intelligent agents}. This resulted in an additional 155 papers, totaling 414 papers. 

The first author reviewed all papers initially by reading the abstract and main contributions. Both authors then discussed and defined the inclusion and exclusion criteria. Given the exponential growth of research in this space and our objective for rigorous analysis, we conservatively focused on papers emphasizing the role of human expertise and tasks in VIS4ML design. To be included in our sample, we required that (1) the paper contribute at least one \textit{interactive visualization} for the purpose of facilitating human analysis of ML, (2) the paper clearly articulated VIS4ML tasks (i.e., how the visualization tool \textit{can effect change} to ML components through a HITL approach), (3) the paper specified one or more \textit{design goals} for the proposed visualization system rationalizing the type of human tasks or activities in the ML pipeline, and (4) the paper included some form of evaluation. 

Consequently, this eliminated a large number of papers that (1) did not state any clear design hypotheses, (2) focused on downstream usage and understanding of trust and fairness, or (3) presented tools designed to stimulate reflection on visualization techniques that could be applied in a machine learning pipeline, but which could also be used for other purposes (e.g.,~\cite{smilkov2016embedding}). After filtering, we had 76 papers in our set. Of these, we sampled 52 papers. Hence, our analysis is not intended to be comprehensive or produce a taxonomy. Instead, we sought to conduct a relatively deep analysis of all of the papers in the sample (see Section~\ref{sec:analysis}) toward assessing the generalizability gaps of VIS4ML research contributions. 

\subsection{Codebook Development}
Initially, both authors independently coded the same set of 5 papers with the high-level goal of identifying 1) how researchers motivated the incorporation of human knowledge in the ML pipeline and what forms this knowledge took, 2) what knowledge outputs or ``insights'' the system intended to help users reach, 3) what sorts of visualizations and automated processes they relied on, and 4) how the design hypotheses implied by the specific motivating claims were evaluated. In this initial read of the papers, we incorporated top-down influences in the form of codes adapted from prior taxonomies (section~\ref{sec:taxonomy}) and bottom-up influences where we identified codes from our observations. 

The authors then discussed the specific aspects of the papers that are important to characterizing threats to the generalizability of VIS4ML research contributions. Through this discussion, we developed our codebook capturing aspirations about \textit{why} the VIS4ML systems were being built, what specific types and examples of insights the papers provide to argue that human knowledge can be extended into the ML pipeline through visualization interfaces, and what datasets and modeling techniques were used in the examples. We also discussed how different sections of the paper map to different codes in our codebook. The codes span across the overall \textit{paper-level} codes such as the paper's motivation and the nature of human expertise required to use the VIS4ML system, and \textit{insight-level} in which the expert inputs their prior knowledge to glean specific insights about the ML components. Our final codebook with descriptions is available in the supplementary materials. Here we describe the main categories using examples from a recent paper on CNNs~\cite{liu2016towards}.

At the paper level, we coded elements like the broader \textit{Motivating Claims for VIS4ML} (e.g., ``[using the tool] experts can diagnose the potential issues of a model and refine a CNN, which enables more rapid iteration and faster convergence in model construction''), how authors \textit{Identified Support Needs for VIS4ML} (e.g., meeting regularly with six deep learning experts over twelve months and including three as authors),
the \textit{Target Generalization Context} for the system (including the properties of the models or datasets it is intended to generalize to, e.g., "CNNs that can be formulated as a DAG with less than 100 classes"), and the overall \textit{Evaluation Approach} (e.g.,  ''visual data analysis and reasoning case study'', adapted from an existing taxonomy for visualization evaluation~\cite{isenberg2013systematic}), as well as the specific \textit{Evaluation Metrics} for tracking whether a system was useful (e.g., ``expert insights''). 

We used the insight level as a more specific unit of analysis to differentiate different types of insights described as supported by the system. We coded \textit{Forms of Human Generated Insights} based on descriptions of knowledge gained about the modeling pipeline (e.g., ``[expert] identified that neurons in the lower layers learned to detect low-level features such as corners\ldots''~\cite{liu2016towards}). Given the prevalence of claims that integrating human knowledge via VIS4ML leads to improved ML use, we coded any \textit{Actions Taken} that authors described resulting from system use. These could be concrete operations applied to the modeling pipeline (e.g.,  the expert added a batch normalization network then retrained, lowering model error by 9\%) to more broadly construed future actions (e.g., the expert suggested they would use the system in their future model development process).

We coded dependencies for each insight, including the \textit{Human Knowledge Required} to reach that insight (e.g., existing domain-specific or domain-general knowledge), the forms of \textit{Feedback via Model Signals} and \textit{Dimensionality Reduction and Other Approximations} such as preprocessing steps or other use of algorithms to transform the data on which the visualizations or interactions depend. We also coded the specific \textit{Dataset} and \textit{Model} associated with the insight (e.g., CIFAR-10 with a 10+2 layer CNN with cross-entropy loss and ReLu~\cite{simonyan2014very}).

\vspace{-1mm}
\subsection{Coding and Analysis Procedure} \label{sec:analysis}
Both authors independently coded a distinct set of 27 and 25 papers (including re-coding the initial five papers). Each paper took between $2-3$ hours to code in which the authors engaged in a \textit{deep reading} of the paper to extract and map the individual information onto the codebook in separate Google Spreadsheets. This included tracing each example insight presented in the paper, identifying the human knowledge that went into generating the insight, specific configurations, and encodings of the visualization, and interaction parameters. Throughout the coding process, the authors also made notes about salient observations about VIS4ML contributions. After coding all the papers, the authors collaboratively analyzed the coded data within each category of codes. This included weekly hour-long discussions and using digital affinity diagramming on Microsoft Whiteboard to cluster the data within each column. The authors also participated in two 3-hour long co-located discussion sessions to synthesize findings about gaps in generalizability and brainstormed recommendations for addressing them. 

Naturally, our analysis is influenced by our research backgrounds. Both researchers have extensive experience in visualization research, which between them includes prior work covering aspects of collaborative design and development of human-centered AI, evaluation practices in visualization (including ML), research transparency, and statistical decision theory. Neither author identifies as an ML researcher.

\begin{table*}[t!]
\fontsize{7}{5}\selectfont
\centering
   \begin{tblr}{  
        colspec={ll|lll|lllllll|lllllll|cc|l|l|lllllll},
        colsep = 1.35pt,
        row{1}={font=\bfseries},              
        row{even}={bg=gray!8},
        row{2} = {font=\bfseries, bg=white},
   }
       & & \SetCell[c=3]{c}{{{\textcolor[HTML]{488A99}{Experts}}}} & & & \SetCell[c=7]{c}{{{\textcolor[HTML]{488A99}{Prior Knowledge}}}} & & & & & & & \SetCell[c=7]{c}{{{\textcolor[HTML]{BA9044}{HITL Task}}}} & & & & & & &  \SetCell[c=2]{c}{{{\textcolor[HTML]{BA9044}{Action}}}} & & \SetCell[c=2]{c}{{{\textcolor[HTML]{4D585B}{ML}}}} & & \SetCell[c=7]{c}{{{\textcolor[HTML]{6AB187}{Evaluation}}}} & & & & & & &  \\
       \# & System & \textcolor[HTML]{488A99}{MLE} & \textcolor[HTML]{488A99}{DoE} & \textcolor[HTML]{488A99}{DAE} & \textcolor[HTML]{488A99}{Dt} & \textcolor[HTML]{488A99}{Do} & \textcolor[HTML]{488A99}{ML} & \textcolor[HTML]{488A99}{C} & \textcolor[HTML]{488A99}{SR} & \textcolor[HTML]{488A99}{Dg} & \textcolor[HTML]{488A99}{T} & \textcolor[HTML]{BA9044}{Tr} & \textcolor[HTML]{BA9044}{I} & \textcolor[HTML]{BA9044}{E} & \textcolor[HTML]{BA9044}{FE} & \textcolor[HTML]{BA9044}{MS} & \textcolor[HTML]{BA9044}{MD} & \textcolor[HTML]{BA9044}{D} & \textcolor[HTML]{BA9044}{O} & \textcolor[HTML]{BA9044}{H} & \textcolor[HTML]{4D585B}{Model} & \textcolor[HTML]{4D585B}{Data} & \textcolor[HTML]{6AB187}{VDAR} & \textcolor[HTML]{6AB187}{UP} & \textcolor[HTML]{6AB187}{AP} & \textcolor[HTML]{6AB187}{UE}  & \textcolor[HTML]{6AB187}{C-QRI} & \textcolor[HTML]{6AB187}{I-QRI} & \textcolor[HTML]{6AB187}{CTV} \\ 
        \toprule
        1 & DataDebugger~\cite{xiang2019interactive} & \greysquare & \greysquare & & \greysquare & & & & & & & & &&&&& \HITLsquare& \HITLsquare&& & MNIST & \vissquare & \vissquare & \vissquare&&&&& \\
        2 & AdViCE~\cite{gomez2021advice} & \greysquare & & & \greysquare & & & & \greysquare & & & & \HITLsquare &&&&&& && SVM & HELOC* &\vissquare&&&&&&&\\
        3 & iForest~\cite{zhao2018iforest} & & & &  & & & & \greysquare & \greysquare & & & \HITLsquare &&&&&& && RF & Titanic, GC&\vissquare&\vissquare&&\vissquare&&&&\\
        4 & INFUSE~\cite{krause2014infuse} & & &\greysquare & \greysquare  & & & & & & & & &&\HITLsquare&&&& &\HITLsquare&LR,DT,kNN &Diabetes&\vissquare&&&&&&&\\
        5 & ActiVis~\cite{kahng2017cti} & \greysquare & \greysquare & & \greysquare & & & & & & & & \HITLsquare &&&&&& &\HITLsquare& CNN&TREC& \vissquare&&&\vissquare&&&&\\
        6 & REMAP~\cite{cashman2019ablate} & \greysquare & & & & & \greysquare & \greysquare & & \greysquare & \greysquare & &&&&\HITLsquare&&& \HITLsquare&& DNN&CIFAR-10&\vissquare&&&\vissquare&&&&\\
        7 & RetainVis~\cite{kwon2018retainvis} & \greysquare & \greysquare & & \greysquare & \greysquare & & & \greysquare & && &&\HITLsquare&&&&& \HITLsquare&\HITLsquare& RNN& HIRA-NPS&\vissquare&&\vissquare&\vissquare&&\vissquare&&\\
        8 & RuleMatrix~\cite{ming2018rulematrix} & & & & & \greysquare & & & & & & & \HITLsquare&&&&&& \HITLsquare&& NN & PIMA&  \vissquare&\vissquare&\vissquare&\vissquare&&&&\\
        9 & DeepVID~\cite{wang2019deepvid} & \greysquare & & & & & & & & & & & \HITLsquare&&&&&& && CNN & MNIST& \vissquare&&&\vissquare&&\vissquare&& \\
        10 & BaobabView~\cite{van2011baobabview} & & & & & \greysquare & & & & & & & \HITLsquare&&\HITLsquare&&&& \HITLsquare&& DT& Oncology&\vissquare&&&&&&&\\
        11 & Perturber~\cite{sietzen2021interactive} & & & & \greysquare & & & & \greysquare & & & &&\HITLsquare&&&&& && CNN & ImageNet&\vissquare&&&&\vissquare&&&\\
        12 & VisLRPDesigner~\cite{huang2021visual} & & & & & & & & & & & &\HITLsquare&&&&&& && CNN & ImageNet&\vissquare&&&\vissquare&&&&\\
        13 & TopoAct~\cite{rathore2021topoact} & & & & & & & & & & & & \HITLsquare&&&&&& &&DNN&ImageNet&\vissquare&&\vissquare&&\vissquare&\vissquare&&\\
        14 & Visevol~\cite{chatzimparmpas2021visevol} & & & & & & \greysquare & \greysquare & & & & & & & &  \HITLsquare & \HITLsquare & & \HITLsquare&&&Bio*&\vissquare&&&\vissquare&&\vissquare&&\\
        15 & Boxer~\cite{gleicher2020boxer} & & & \greysquare & & & & & & & & & &&&\HITLsquare&&& \HITLsquare&&&IMDB&\vissquare\\
        16 & DeepEyes~\cite{pezzotti2017deepeyes} & \greysquare & & & & & & & & & & &&&&&\HITLsquare&& \HITLsquare&&CNN&MNIST&\vissquare&&&&&\vissquare&&\\
        17 & Blocks~\cite{bilal2017convolutional} & & & & & & & & & & & & \HITLsquare&&&&&& \HITLsquare&\HITLsquare&CNN&ImageNet&\vissquare&&&\vissquare&&&&\\
        18 & DQNViz~\cite{wang2018dqnviz} & \greysquare & & & & & & \greysquare & & & & \HITLsquare & & & & & & & \HITLsquare&\HITLsquare& DQN&&\vissquare&&&\vissquare&&&&\\
        19 & ConceptExplainer~\cite{huang2022conceptexplainer} & & & & & & & & & & &  & \HITLsquare&&&&&& && CNN&ImageNet&\vissquare&&&\vissquare&&&&\\ 
        20 & SliceTeller~\cite{zhang2022sliceteller} & \greysquare & & & \greysquare & \greysquare & & & & & \greysquare & &&\HITLsquare&&&&& \HITLsquare&\HITLsquare&DNN&&\vissquare&&&\vissquare&&&&\\ 
        21 & NAS-Navigator~\cite{tyagi1912navigator} & & & & & & & & & &  \greysquare & &&&&\HITLsquare&&& \HITLsquare&&CNN&CIFAR-10*&\vissquare&&\vissquare&\vissquare&&&&\\
        22 & FSLDiagnotor~\cite{yang2022diagnosing} & \greysquare & & & & & & & & & & &&&&\HITLsquare&&& \HITLsquare&&CNN&ImageNet&\vissquare&&\vissquare&\vissquare&&&&\\ 
        23 & FeatureEnVi~\cite{chatzimparmpas2022featureenvi} & & & & \greysquare & \greysquare & \greysquare & & & & & & &&\HITLsquare&&&& \HITLsquare&& XGBoost&UCI&\vissquare&&&\vissquare&&&&\\ 
        24 & GNNLens~\cite{jin2022gnnlens} & & & & &&&&& \greysquare & & &&\HITLsquare&&&&& &\HITLsquare& GNN&Cora-ML&\vissquare&&&\vissquare&&&&\\ 
        25 & HetVis~\cite{wang2022hetvis} & \greysquare & & & && \greysquare & && \greysquare& &&&\HITLsquare&&&&& &\HITLsquare& CNN & Face Mask&\vissquare&&&\vissquare&&&&\\ 
        26 & DECE~\cite{cheng2020dece} & & & & \greysquare & &&\greysquare&\greysquare&\greysquare&& &\HITLsquare&&&&&& &\HITLsquare& NN & Pima, GC&\vissquare&&\vissquare&\vissquare&&&&\\ 
        27 & NeuroCartography~\cite{park2021neurocartography}  & & & & & & & & & & & & \HITLsquare&&&&&& && CNN&ImageNet&\vissquare&\vissquare&&\vissquare&&&&\\ 
        28 & What-If~\cite{wexler2019if} & & & & &&&&\greysquare&&& & \HITLsquare &&&&&& \HITLsquare&&LR&UCI&\vissquare&&&\vissquare&&\vissquare&&\\ 
        29 & Errudite~\cite{wu2019errudite} & & & & &&\greysquare&&\greysquare&& &&&\HITLsquare&&&&& &&BiDAF&SQuAD&\vissquare&\vissquare&&\vissquare&&&&\\ 
        30 & HardVis~\cite{chatzimparmpas2022hardvis} & & & & \greysquare &\greysquare &&&&&& &&\HITLsquare&&&&\HITLsquare& \HITLsquare&&kNN&Cancer*&\vissquare&&&\vissquare&&\vissquare&&\\ 
        31 & VATUN~\cite{park2021vatun} & \greysquare & & & &\greysquare&&&&\greysquare&& &\HITLsquare&&&&&& &\HITLsquare& CNN & CIFAR-10&\vissquare&&&\vissquare&&\vissquare&&\\ 
        32 & CNN Explainer~\cite{wang2020cnn} & \greysquare & & & & & & & & & & & \HITLsquare&&&&&& &\HITLsquare&CNN&CIFAR-10&\vissquare&&&\vissquare&&&\vissquare\\ 
        33 & LSTM Vis~\cite{strobelt2017lstmvis} & \greysquare & \greysquare & & && \greysquare&&&\greysquare&&& \HITLsquare&&&&&& &\HITLsquare&LSTM&Synthetic&\vissquare&&&\vissquare&&&&\\ 
        34 & SEQ2SEQ-VIS~\cite{strobelt2018s} & \greysquare & & & &&&&\greysquare&&&&&\HITLsquare&&&&& 
 \HITLsquare&\HITLsquare&German-Eng&IWSLT'14&\vissquare&&&&&&&\\ 
        35 & SUMMIT~\cite{hohman2019s} & & & & & & & & & & & &\HITLsquare&&&&&& \HITLsquare&\HITLsquare&CNN&ImageNet&\vissquare&&\vissquare&&&&&&\\ 
        36 & Confusion Wheel~\cite{alsallakh2014visual} & & & & & & & & & & & && \HITLsquare&&&&& &\HITLsquare&&UCI,MNIST&\vissquare&&&\vissquare&&\vissquare&&\\ 
        37 & TensorFlow Graph~\cite{wongsuphasawat2017visualizing} & \greysquare & \greysquare & &  & & & & & & & & \HITLsquare&&&&&& &\HITLsquare& CNN&CIFAR-10*&\vissquare&&&\vissquare&&&&\\ 
        38 & Semantic Navigator~\cite{jia2021towards} & & & & \greysquare&\greysquare&&&&&& &&\HITLsquare&&&&& \HITLsquare&&DNN&&\vissquare&\vissquare&&\vissquare&&\vissquare&&\\ 
        39 & CNNVis~\cite{liu2016towards} & \greysquare & & & &&&&&\greysquare&& &&\HITLsquare&&&&& \HITLsquare&&CNN&CIFAR-10&\vissquare&&&&&\vissquare&&\\ 
        40 & VA Workspace~\cite{el2018visual} & & & & && \greysquare& \greysquare& \greysquare&&& && \HITLsquare&&&&& \HITLsquare&&&&\vissquare&\vissquare&\vissquare&\vissquare&&&&\\ 
        41 & DGMTracker~\cite{liu2017analyzing} & \greysquare & & & && \greysquare&&&&& \HITLsquare&&&&&&& &\HITLsquare&GAN&CIFAR-10&\vissquare&&&&\vissquare&&&\\ 
        42 & AEVis~\cite{liu2018analyzing} & \greysquare & & & \greysquare&&&&&\greysquare&&&&\HITLsquare&&&&& &&CNN&ImageNet&\vissquare&&&&\vissquare&&&\\ 
        43 & RNNVis~\cite{ming2017understanding} & \greysquare & & & & & & & & & & & \HITLsquare&&&&&& \HITLsquare&&RNN&Yelp&\vissquare&&&\vissquare&&\vissquare&&\\ 
        44 & TNNVis~\cite{nie2018visualizing} & \greysquare & & & \greysquare &&\greysquare&&&&& & \HITLsquare&&&&&& &\HITLsquare&PoS&&\vissquare&&&&&\vissquare&&\\ 
        45 & SCS~\cite{el2019semantic} & & & & & & & & & & & &&\HITLsquare&&&&& \HITLsquare&&Topic Model&&\vissquare&\vissquare&&\vissquare&&&&\\ 
        46 & BOOSTVis~\cite{liu2017visual} & \greysquare & & & &&\greysquare&&&&&\HITLsquare&&&&&&& \HITLsquare&&GBDT&&\vissquare&&&&&\vissquare&&\\ 
        47 & SCANViz~\cite{wang2020scanviz} & \greysquare & & & & & & & & & & &\HITLsquare&&&&&&\HITLsquare&\HITLsquare&SCANViz*&&\vissquare&&&\vissquare&&&&\\ 
        48 & DRLIVE~\cite{wang2021visual} & \greysquare & & & & & & & & & & & \HITLsquare&&&&&& \HITLsquare&\HITLsquare&DRL&Atari&\vissquare&&&\vissquare&&\vissquare&&\\ 
        49 & ProtoSteer~\cite{ming2019protosteer} & \greysquare & \greysquare & & \greysquare &&&&\greysquare&&&&&\HITLsquare&&&&& \HITLsquare&\HITLsquare&ProSeNet&Yelp&\vissquare&&&&\vissquare&&\\ 
        50 & GAN Lab~\cite{kahng2018gan} & & & & &&&&&&\greysquare&&\HITLsquare&&&&&& &\HITLsquare&GAN&&\vissquare&&&&&\vissquare&&\\ 
        51 & Beames~\cite{das2019beames} & & & & & & & & & & & &\HITLsquare&&&&&& &\HITLsquare&LR&Housing&\vissquare&&&&&&&\\ 
        52 & DRLViz~\cite{jaunet2020drlviz}  & \greysquare & & & & & & & & & & &\HITLsquare&&&&&& \HITLsquare&\HITLsquare&DRL&&\vissquare&&&\vissquare&&&&\\                
        \bottomrule
   \end{tblr}
   \caption{List of VIS4ML papers and key paper-level columns considered in our analysis. \textbf{Expert} engagement includes Machine Learning Experts \textbf{(MLE)}, Domain Experts \textbf{(DoE)}, and Data Analysts \textbf{(DAE)}. Assumed expert \textbf{prior knowledge} includes: Domain Knowledge \textbf{(Do)}, Data Knowledge \textbf{(Dt)}, Machine Learning Knowledge \textbf{(ML)}, Tacit Knowledge \textbf{(T)}, Scientific Reasoning ability \textbf{(SR)}, Choice Assessment \textbf{(C)}, and Diagnostics skills \textbf{(Dg)}. The \textbf{human-in-the-loop tasks} are Interpreting and Assessing models \textbf{(I)}, Model Selection/Choice\textbf{ (MS)}, Debug and Fix Errors \textbf{(E)}, Model Design \textbf{(MD)}, Model training \textbf{(Tr)}, Feature Engineering\textbf{ (FE)}, and Examining/Preparing Data \textbf{(D)}. \textbf{Insight-informed actions} include Observed Actions \textbf{(O)} and Hypothetical Actions \textbf{(H)}. \textbf{Evaluation} taxonomy is based on ~\cite{isenberg2013systematic}}.
    \label{tab:my_label}
\end{table*}

%% file: 04_results.tex
\section{Findings}

We organize our findings based on how the 52 papers in our sample characterize humans involved in VIS4ML, the scope of the ML components and pipelines, the intended HITL tasks supported with VIS4ML, and the approach to implementing VIS4ML tools given existing workflows and evaluation of the overall system and critical abstractions. 

\subsection{Characterizing Humans in VIS4ML}
Ideally, to innovate VIS4ML tools, researchers should determine the specific nature of human expertise, prior knowledge, and skills \textit{representative} of real-world workflows. While a majority of papers (76.9\%) directly engaged with experts -- individuals who have the necessary expertise to intervene in the ML modeling process -- to identify requirements for VIS4ML, many lacked rationale for why and how those experts were sampled and the nature of their expertise in supporting HITL tasks. 

To involve stakeholders, researchers employed various need-finding methods, including interviews (10; 19.2\%), regular meetings with experts (14;26.9\%), iterative design and evaluation (11, 21.1\%), and participatory design (3; 5.7\%). Across these approaches, participants comprised ML experts \textcolor[HTML]{488A99}{[MLE]} (26; 50\% ), domain experts \textcolor[HTML]{488A99}{[DoE]} (5; 9.6\%), or data analysts \textcolor[HTML]{488A99}{[DAE]}  (2; 3.8\%). Further, only five papers (9.6\%) included both ML and domain experts as study participants.  Considering that many of the VIS4ML systems require prior knowledge or expertise spanning across ML and specific domains such as health, multi-stakeholder involvement is not prevalent. The number of participants in formative studies ranged from 2 - 20, and in 3 papers, the authors themselves were expert participants. While papers often implied that engaging with experts was critical to ensure the validity of their work, very few systematically reported details about the study protocol, including descriptions of specific expertise or expert knowledge, recruitment and study design, or the nature of design probes and feedback, so as to enable reproducing the methods. 

\subsubsection{Assumptions about Prior Knowledge and Skills}
Almost all papers lacked explicit descriptions of prior knowledge and skills required of human experts to inform VIS4ML designs. However, authors' descriptions of design goals based on formative studies surfaced \textit{implicit} assumptions and prerequisites about the expertise and skills required to be in the loop. Of the 52 papers, 11 (21.1\%) mention the need for \textbf{machine learning knowledge \textcolor[HTML]{488A99}{[ML]}}, including the conceptual understanding of specific models and modeling techniques, practical (hands-on) experience with training models, and the ability to comprehend model statistics and performance. For example, \textit{LSTM Vis}~\cite{strobelt2017lstmvis} assumes that \textit{``architects are deeply knowledgeable about machine learning, neural networks, and the internal structure of the system''} Further, 14 papers (26.9\%) emphasize \textbf{data knowledge \textcolor[HTML]{488A99}{[Dt]}} or expertise required to use VIS4ML systems. Data knowledge includes familiarity with specific datasets, contextual understanding of data and sub-groups, ground truth labels, and the ability to judge the relative importance of data instances and classes. \textit{HardVis}~\cite{chatzimparmpas2022hardvis} assumes that users are \textit{``competent in judging the influence of a suggestion on the whole data set''} when exploring automated sampling suggestions.

In 8 papers (15.3\%), authors indicated the need for \textbf{domain knowledge \textcolor[HTML]{488A99}{[Do]}}, including domain-general knowledge (e.g., how language works), the ability to contextualize and comprehend model decisions, the ability to recognize good and bad model behavior, and the ability to notice errors and foresee domain consequences of bad model behavior. In describing the task of analyzing feature transformation, \textit{FeatureEnVi}~\cite{chatzimparmpas2022featureenvi} requires that \textit{``A user should be competent in judging the influence of feature transformations before applying them.''} Lastly, 4 papers (7.7\%) based their design choices on experiential or \textbf{tacit knowledge \textcolor[HTML]{488A99}{[T]}} about modeling pipelines, including prior knowledge about performant network architecture, known constraints to search the architecture space, past experience revising the space of hyper-parameters, and knowledge about critical data examples to assess model behavior. 

Papers also made assumptions about the specific skill sets required to interact with VIS4ML systems. From our analysis, we identified three types of skill sets that are a combination of domain, data, and ML knowledge. Ten papers (19.2\%) assume that users are able to perform \textbf{diagnostic analysis \textcolor[HTML]{488A99}{[Dg]}} of the modeling pipeline, including running ablation studies, analyzing model behavior using adversarial examples, and root cause identification through exploration and inspection. Nine papers (17.3\%) require that users are able to engage in \textbf{scientific reasoning \textcolor[HTML]{488A99}{[SR]}} through hypothesis generation and testing, counterfactual analysis, and case-based reasoning. Finally, the design paradigms for four papers (7\%) are based on choice architecture, i.e., the ability to\textbf{ assess choices \textcolor[HTML]{488A99}{[C]}} and make modeling decisions (e.g., comparing models, selecting from a list of automated recommendations, etc.).

\subsection{Scope of ML Components and Modeling Pipeline}
Based on need-finding studies with experts, papers defined concrete task requirements (36; 69.2\%), identified design challenges (7; 13.4\%), and derived design goals (20; 38.4\%) for VIS4ML systems. By analyzing these requirements across papers, we identified researchers' \textit{aspirations} about the scope of VIS4ML contributions within the modeling pipeline. 
Papers aspired to address critical challenges of \textbf{scale, generalizability, high dimensionality, perceptibility, and varied data types} in the ML pipeline. For instance, \textit{ConceptExplainer}~\cite{huang2022conceptexplainer} aims to support multi-scale concept visualization (e.g., ImageNet dataset with 1.2 M images for 1000 classes), \textit{REMAP}~\cite{cashman2019ablate} aims to lower time and resource cost of finding performant model architectures, and \textit{ActiVis}~\cite{kahng2017cti} seeks to solve model exploration for multiple types of data. Other systems aim to help analysts perceive and discover salient patterns and insights in high-dimensional data and complex model architectures (e.g.,~\cite{hohman2019s}).

\subsubsection{Datasets and Model Types in VIS4ML Implementation}
All but two papers in our sample report on specific data and models used to construct examples in the paper or run an evaluation. Most papers relied on established benchmark datasets to demonstrate tools, including but not limited to ImageNet (9; 17\%), CIFAR-10 (7; 13\%), MNIST (6; 11.5\%), Yelp restaurant reviews (3; 6\%), and other examples from the UCI ML repository (6; 11.5\%). Several demonstrated the tool using synthetic data (3; 6\%). We observed some disparity in how papers valued using simple examples and well-known datasets. Many implied that using well-known datasets or benchmarks made their work stronger, but some papers commented on the need for tools to support and be evaluated on data that experts care about, for example, because using popular data like MNIST for better verification led to little insight about diagnosis and model refinement due to its simplicity~\cite{liu2016towards}. Others constructed systems, for example, for pedagogical purposes, that were acknowledged to be unrealistic (e.g., simple 2D data) to avoid dependencies on approximations like dimensionality reduction~\cite{kahng2018gan} in surfacing the results for users.  

In terms of target model types, many systems are designed for convolutional neural nets (CNNs: 20; 38.5\%). Others focused on recurrent neural nets (RNNs: 8; 15.4\%), generic deep neural networks (DNNs: 3, 5.8\%), decision tree-based approaches (7, 13.5\%), k-nearest-neighbors (3; 5.8\%), GANs and deep reinforcement learning (2 each), and zero-shot models, graph neural networks, and deep Q networks (1 each). Many systems were framed as intended for large models or datasets; however, scalability in terms of classes or concepts was often a stated limitation. Scalability constraints were described in terms of numbers of classes (e.g., up to 20, 100, 1000, etc.), concepts (up to 40, 100), features, instances, feature maps, and nodes in a max-pooling layer, as well as dimensionality (of both datasets and hidden states). Some systems were described as intended for small models (e.g.,~\cite{cashman2019ablate,wang2020cnn}). Others were highly specific (e.g., GANs that can run in a browser with 2D data samples~\cite{kahng2018gan}). Papers also occasionally described expected performance under other properties of inputs or outputs, like imbalanced data~\cite{alsallakh2014visual} or non-orthogonal concepts~\cite{wang2020scanviz}. 

The papers often commented on the scalability of their contributions to broader classes or data conditions. A few systems were described as directly applicable to varying architectures (e.g.,~\cite{wang2020scanviz}), data modalities like images or text (e.g.,~\cite{chaudhuri2006data}), model types (e.g.,~\cite{alsallakh2014visual,rathore2021topoact}), and encoder types~\cite{strobelt2018s}). However, other times papers made generalizability claims implying that the general combination of representations or interactivity would be adaptable to other cases given further engineering.

\subsection{Human-in-the-loop Tasks}
All papers motivated the need for incorporating human knowledge in the modeling pipeline. By coding the descriptions in the introduction section, we identified seven categories of \textit{human-ML interchange} in the modeling pipeline. While several papers ascribe human roles in multiple stages of the pipeline, 25 papers (48.1\%) specifically emphasize human involvement to \textbf{interpret and assess \textcolor[HTML]{BA9044}{[I]}} model behavior. Further, 15 papers (28.8\%) motivate humans' role in \textbf{debugging and fixing model errors \textcolor[HTML]{BA9044}{[E]}}. In targeting the earlier stages of the modeling pipeline, two papers (3.8\%) mention human inputs in \textbf{examining and preparing data \textcolor[HTML]{BA9044}{[D]}}, three papers (5.8\%) focus on \textbf{feature engineering \textcolor[HTML]{BA9044}{[FE]}}, five papers (9.6\%) highlight the need for human expertise in \textbf{choosing the right modeling techniques \textcolor[HTML]{BA9044}{[MS]}}, two (3.8\%) on \textbf{model design and configuration \textcolor[HTML]{BA9044}{[MD]}}, and three (5.8\%) motivate humans' role in \textbf{monitoring and managing the model training \textcolor[HTML]{BA9044}{[Tr]}} process. 

In identifying these human roles, a majority of papers present motivating claims that fall under one of five categories, including (1) current limitations of machine learning use in real-world, (2) requirements for machine learning applications to succeed, (3) modeling complexity, (4) limitations of current approaches for HITL, and (5) lack of support for incorporating human knowledge in the modeling pipeline. Domain criticality of ML models and the need for trust (e.g., \textit{``knowing how entire classes are represented inside of a model is important for trusting a model's predictions\ldots''}~\cite{hohman2019s}), human effort and experience (e.g., \textit{``\ldots can't blindly trust automated methods (e.g., in a medical setting, doctors will want explanations of predictions),''}~\cite{cheng2020dece}), and scalability are prominent themes across motivation claims, though formal definitions of these goals are not given.

While the majority of reported VIS4ML task requirements are concrete, such as making \textit{comparisons} between classes or models, in a few cases, tasks are defined only in the abstract and lack clear descriptions of scope or task resolution (e.g., `exploring' details or `understanding' model behavior). Further, while many papers ground their design goals in measures of model performance, effort, effectiveness, scale, and heterogeneity of modeling characteristics, their definitions of these measures and whether or how the properties emerged from need-finding studies or were chosen a priori are not always clearly specified. 

To understand ways in which VIS4ML systems support the HITL tasks described above, we coded insight examples provided in usage scenarios or case studies and actions they purportedly inspire.

\subsubsection{Forms of Human Generated Insights}\label{sec:insights}
We identified six categories of insights across visualizations of training data, fitted models, and model representations and configuration. Given that interpretability is a central topic (nearly half of the papers in our sample),  VIS4ML systems are meant to support insight generation about \textbf{inference mechanisms} (i.e., what the model learns and how it makes inferences). Examples and descriptions suggested analysts could generate causal hypotheses about the influence of structure and features on prediction results, how different models learn features, mapping between layers, classes, and concepts, what the model has learned from data, agent strategies in reinforcement learning, and how data attributes such as pixels in images contribute to classification. For instance, at the data level, in~\cite{huang2021visual}, the analyst attempts to learn through pixel-flipping (setting selected pixel values to zero) the relevance of water surface pixels contributing to the `boat house' class. Or, by observing cross-class links of concepts for different vehicle classes (e.g., car windows), the analyst might conclude that the neural network has learned common features across different cars~\cite{huang2022conceptexplainer}.

Further, through exploratory analysis, analysts might gather \textbf{evidence of erroneous behavior} and \textbf{root causes of errors} (i.e., finding and fixing errors). VIS4ML solutions were aimed at surfacing inconsistencies in model decisions, clusters with low accuracy, evidence of concept incoherent topics, lack of clear class separation, edge cases and hard-to-distinguish classes, model vulnerabilities to adversarial perturbations, etc. As an example, in the scatterplot visualization in \textit{DataDebugger}~\cite{xiang2019interactive}, the analyst sees that the classes `knitwear' and `sweater' were heavily mixed. In DeepEyes~\cite{pezzotti2017deepeyes}, the analyst sees that activation of a “digit-5” associated filter also showed strong activation on digit-3, indicating perfect class separation is impossible in that layer. Visualizations also help in debugging and identifying root causes of errors, including denigrated or oversized filters, limitations with neuron cluster composition, concept entanglement, problematic layers in the network, sub-optimal agent strategies, and errors originating in different model components. For example, in DQNViz~\cite{wang2018dqnviz}, the analyst is said to observe that the agent moving the paddle left and right (a strategy in the Breakout game) comprised 31\% and 47\% of 25,000 steps in the epoch but did not contribute to achieving rewards. 

In addition to debugging model and data errors, VIS4ML systems aim to support \textbf{validation} or assertions of intended model behavior. Across our set of papers, analysts were described as being able to evaluate hypotheses about known characteristics of good training (e.g., important classifiers centered at the beginning) and similarities between neural networks and human decision rationales and confirm consistency in the model's decision-making. For instance, in \textit{SUMMIT}~\cite{hohman2019s}, the analyst verifies that, similar to humans, the model classifies black bear and brown bear based on color. Further, VIS4ML systems aim to help analysts understand the \textbf{space of data and modeling choices} for subsequent decisions and actions. Concretely, analysts are thought to gain insights into the strengths and limitations of different feature selection algorithms, feature importance and which features to exclude, comparative differences between models, architecture choices such as which layer to remove in a neural network, which models to include in an ensemble, and how to slice datasets. 

In the process of deducing these different insights, analysts also need to make inferences about \textbf{model performance}. To facilitate this understanding, VA tools present information about the model depth and classification accuracy, model accuracy for different classes, which model performs best, data slices and performance, data quality and ground truth impact on performance, model convergence time, memory, and compute time during training, etc. 

\subsubsection{Model Signals for Insights}
To support users' inference processes, designers of VIS4ML systems must identify what forms of feedback on model quality, interpretability, or performance to surface. Most papers surfaced some form of metric to provide feedback. Of the 52 papers, 23 (44\%) provided information about model performance using confusion matrices, plain text, or line charts showing loss and accuracy curves. In addition, 7 papers (13.4\%) provided information about class probabilities using text or color-coded bar charts. Lastly, 11 papers (21.1\%) provided feedback about modeling tasks such as training time, number of parameters, number of data items corrected through active learning, delta-changes or improvement to model performance, and feature importance. The level of motivation for the specific model signals that papers used varied considerably, with many papers providing a very brief motivation and a few providing more rigorous motivation of why the chosen signals were good estimators for improving model performance.

\subsubsection{Insight-informed Actions}
\label{sec:actions}
 Given the prevalence of claims that integrating human knowledge via VIS4ML leads to improved ML use, we wanted to see what sorts of actions---from concrete operations applied to the modeling pipeline to more broadly construed future actions---papers described resulting from the insight gains of a system. For each human-generated insight, we coded any actions described as taken based on that insight. 
 
We noted whether the described actions occurred in the context of \textit{author-provided} examples, such as fictional case studies or running examples used throughout the paper, or \textit{expert case studies}. We also noted whether the actions were \textbf{observed \textcolor[HTML]{BA9044}{[O]}} (e.g., actually applied to a modeling pipeline), such as when papers described actions an expert took based on using the tool or presented results from re-training a model after implementing a change, versus actions that were \textbf{hypothetical \textcolor[HTML]{BA9044}{[H]}}, i.e., that could be taken or were referred to as possible future steps. Finally, we noted whether specific before and after performance comparisons were made (e.g., accuracy comparisons) to validate the utility of observed actions in the larger ML pipeline or research endeavor.

\textit{Observed actions:}
Overall, the majority of papers (42; 82.7\%) described some action as the result of an insight gained from the VIS4ML system. Of these, 27 (about half of the total 52 papers) described at least one observed action. About a third of the total papers (15 of 52) described at least one action taken by an expert. The remainder (23.1\% of 52) were actions by the authors as part of the case studies they designed. 
Examples of actions that authors or system users took based on insights include
\textit{changes to the training data}, such as sampling to deal with class imbalances or modifying training labels; 
\textit{changes to the features}, 
like switching input images to grayscale~\cite{bilal2017convolutional};
\textit{changes to a model representation}, 
like adding or removing rules from rule-based classifiers~\cite{alsallakh2014visual} or moving neurons between clusters~\cite{liu2016towards}; \textit{changes to model parameters}, like reducing the number of latent dimensions~\cite{wang2020scanviz}; changes to the model architecture, like adjusting layers or filters of deep NNs~\cite{cashman2019ablate,pezzotti2017deepeyes}; 
and changes to the training process, like changing the learning rate~\cite{kahng2018gan} or variance sampler~\cite{liu2017analyzing} of deep generative models.

Slightly more than one third of the total papers (19 of 52) described or quantified the effect of an action on the modeling pipeline. Most cited numeric changes in model accuracy or error. Several papers presented accuracy, precision, and recall statistics (e.g.,~\cite{chatzimparmpas2022featureenvi}) to acknowledge trade-offs or changes in training speed (e.g.,~\cite{pezzotti2017deepeyes}). A few other papers reported on trade-offs between accuracy and interpretability, such as how accuracy remained similar after human-driven adjustments while the number of nodes in the latent representation significantly reduced~\cite{van2011baobabview}.
Hence, over 60\% of the papers in our sample provided no concrete evidence of the impacts of the VIS4ML system on the ML pipeline.

\textit{Hypothetical actions:}
Of the 25 papers where no concrete action was observed, 15 (28.5\%) described a hypothetical action. Six papers (11.5\%) referred to hypothetical actions proposed by an expert in a case study; the remainder were proposed by authors as part of examples they developed.
Hypothetical actions could be well defined, such as when papers mentioned specific actions on a pipeline that could be taken upon reaching some specific insight, e.g., reparameterizing a deep RL system after observing that a lower dimensional representation appears to have explanatory power~\cite{wang2021visual}. Other hypothetical actions on a pipeline were referred to in less specific terms, such as informing data collection or experiment design~\cite{bilal2017convolutional,ming2019protosteer,strobelt2017lstmvis}, allowing fine-tuning of a model~\cite{alsallakh2014visual,wang2021visual,wexler2019if}, or supporting bug finding~\cite{wexler2019if}.

Other hypothetical actions concerned changes to experts' processes, such as when experts said that they would incorporate the system in their future model development processes~\cite{liu2016towards}, or their insights were thought to inform subsequent analysis or theoretical investigations~\cite{ming2017understanding,strobelt2017lstmvis,wang2018dqnviz}, or future research or other "endeavors" to improve such models~\cite{cheng2020dece,jin2022gnnlens}. 

Of the 10 papers where neither observed nor hypothetical actions were described, one described a system developed for educating non-experts~\cite{wang2020cnn}, noting that they chose this goal because supporting an interactive training process would be unrealistic. Other papers in this group focused on interpretation goals without necessarily providing reasons why actions were not considered.

\subsection{VIS4ML Implementation and Evaluation}
Based on need-finding studies, papers identified implementation desiderata for supporting human understanding, design, and improvement of the model. Specifically, the papers aimed for their systems to \textbf{align with existing modeling practices and workflows.} Through their interactions with experts, authors either captured existing task workflows or defined new workflows for VA tasks. For instance, \textit{FeatureEnVi}~\cite{chatzimparmpas2022featureenvi} defines a unified workflow for feature engineering by ``fusing stepwise selection and semi-automatic extraction approaches.'' A few papers emphasized \textbf{human-knowledge integration and guided discovery} of model and data characteristics. In rationalizing system design considerations, authors made connections to human knowledge and comprehension support needs. For instance, in designing \textit{NAS-Navigator}~\cite{tyagi2022navigator}, authors intended that users be able to design and edit template models based on their experience. Additionally, some authors hoped to \textbf{support varied work environments} and overcome ML deployment challenges. 

Prior taxonomies broadly describe the specific visualizations and interaction formats we observed across papers. Hence we focus on papers' reliance on abstraction techniques for model interpretability and approaches to VIS4ML evaluation. 

\subsubsection{Post-Hoc Interpretability Methods}
\label{sec:dimr}
We observed frequent use of dynamic algorithms and approximating representations in the visualizations employed in VIS4ML tools. Many papers used dimensionality reduction techniques (e.g., PCA, UMAP, MDS; 12 papers; 23.1\%) and/or projection-based visualization techniques like t-SNE (16; 30.8\%) to visualize high dimensional data in 2D. While t-SNE generated layouts are guaranteed to recover structure in high dimensional data under certain conditions~\cite{arora2018analysis}, without proper tuning, they can produce artifacts that mislead users to perceive structure that doesn't exist~\cite{wattenberg2016use}. 
Two papers seemed to acknowledge such limitations, noting that they had intentionally opted not to use projections like t-SNE, for example, because the authors ``found that if the tool loses its inherent connection to the data, results were less interpretable to the user''~\cite{strobelt2017lstmvis}. 

Similarly, the machine learning interpretability literature has contributed a number of post-hoc explanation methods designed to provide intuition into how a model reaches a decision or what it has learned. The VIS4ML systems we surveyed made frequent use of feature attribution approaches such as feature visualization through partial dependency plots and other graphics in feature space (13; 25\%) or deriving of importance scores for ranking or recommendation (7; 13.5\%).
 
Pixel-based saliency maps and other forms of activation heatmaps were also used (5; 9.6\%).  
In a few cases, papers referred to limitations of these approaches, such as by discussing how maps of salient image patches activating a neuron were not appropriate for explaining the activity in neurons of very deep CNNs, where activations are influenced by very large patches~\cite{liu2018analyzing}. 

For VIS4ML tools to be implemented in practice requires transparency on how hyperparameters used in critical approximating representations are set, including any tuning processes used. Papers varied considerably in the extent to which they specified such information for dynamic visualization algorithms like t-SNE, clustering approaches, or other optimizations. Interactive hyperparameter tuning, such as the ability to set a target k for clustering algorithms, was occasionally made available to end users as a strategy to avoid dependence on authors' decisions. In cases where it was not but a parameterized technique was used, some authors described only the values they set, while others described the values and the tuning process they used, and others provided suggestions for how those adapting the system should use or not use the specific values or process they used. Occasionally papers did not commit to any single technique, instead mentioning various projection methods that could be used (e.g.,~\cite{strobelt2018s}).

\subsubsection{VIS4ML Evaluation}
To understand VIS4ML evaluation as a subset of visualization evaluation more broadly, we used Isenberg et al.'s~\cite{isenberg2013systematic} adaption of Lam et al.'s taxonomy \cite{lam2011empirical} to characterize forms of visualization evaluation and the metrics or outcome variables associated with them. We observed instances of all evaluation styles except \textit{Evaluating Collaborative Data Analysis}. We report on formative studies aimed at \textit{Understanding Work Practices} described above. 

Most notably, every paper in our sample exemplified validation through \textbf{Visual Data Analysis and Reasoning \textcolor[HTML]{6AB187}{[VDAR]}}. VDAR captures case-study style evaluations of how a visualization tool supports analysis and reasoning about data and allows domain experts to derive knowledge. Case studies could be based on collaboration or observation of expert users or describe hypothetical users' analysis processes (Section~\ref{sec:actions}). Sometimes experts were paired with authors in case studies, though similar to Sperlle et al.'s~\cite{sperrle2021survey} observation of missing details in evaluation, it was not always clear whether authors were involved. In reporting these scenarios, however, papers narrated VDAR processes by describing how observations at various points could be interpreted as evidence of certain facts about the data, model, or pipeline. 

The implied evaluation metric of a VDAR-style evaluation is the validity of the insights that are generated. As discussed above, only about one-third of the papers provided evidence of improved model performance as a result of insight-informed actions. Hence, most of these evaluations inherit the ambiguity of what makes an insight meaningful or important, which is frequently discussed in visualization research~\cite{north2006toward, yi2008understanding}. In a couple of papers, authors performed robustness testing to establish the validity of insights, such as testing to verify that a user's insights generalize beyond the specific input data and/or model architecture used~\cite{sietzen2021interactive} or using PCA to validate structures identified using a contributed topological summary approach for exploring DNN activations~\cite{rathore2021topoact}.  

More commonly, however, expert judgment was implied to establish that insights were useful.  Authors often signaled why a process was meaningful by interspersing quotes from the experts' thinking aloud as they used the tool. For example, some papers included quotes from experts describing how they were able to use a system to confirm prior knowledge, such as confirming evidence of under- or overfitting in comparing CNN models that varied in complexity~\cite{liu2016towards}. In other cases, quotes summarized how the insights or actions they were able to achieve with the system represented solutions to problems they often face (e.g.,~\cite{liu2016towards}). Others described how insights stimulated further consideration on the part of users.

A few papers implied that an insight gained from a VDAR process was valid because it corroborated an observation or a hunch identified in prior work. For example, papers describe how an observation ``confirms earlier work that demonstrates simple context-free models in RNNs and LSTMs''~\cite{strobelt2017lstmvis,strobelt2018s} or how the absence of positive values in a DNN's training process was ``observed, but not explained'' by prior work~\cite{liu2017analyzing}. 

VDAR evaluations were commonly paired with subjective feedback and opinions from users (\textbf{User Experience \textcolor[HTML]{6AB187}{[UE]}}; 65.4\% of total papers). Many papers used think-aloud protocol and/or semi-structured interviews to gather qualitative feedback from VDAR participants. Others elicited Likert-style responses on the usefulness and usability of a tool. Less commonly, surveys were used to elicit experts' self-assessments of their findings (e.g.,~\cite{wu2019errudite}). This was rare, however; in most cases, it was implicit that because an expert made an observation, they must have faith in that observation. User Experience observations were most frequently used to establish that experts found the system useful and to describe potential improvements. 

Sixteen papers (30.8\%) included what prior surveys of visualization evaluations have called \textbf{isolated Qualitative Results Inspection \textcolor[HTML]{6AB187}{[I-QRI]}}, referring to validation of techniques by inspecting a technique's results in isolation, along with a description of how the technique achieves some result. All isolated QRI examples we observed were used to validate the visualization techniques used. For example, papers might point to how a certain pattern exemplified in a visualization was indicative of some fact about the latent representation, training process, or other parts of the pipeline: ``it is evident in Fig. 3b that the digits 1, 2 and 7 are often confused with each other, as their shapes are similar to some degree~\cite{alsallakh2014visual}. On the other hand, five papers (9.6\%) used \textbf{comparative (QRI) \textcolor[HTML]{6AB187}{[C-QRI]}} to compare visual outputs to other state-of-the-art techniques to suggest that an adopted approach is superior (e.g., ``Without Fourier basis parametrization, the differences between the models are more visually distinct''~\cite{sietzen2021interactive}).

Beyond qualitative evaluations of user experience, we observed a smaller number of papers (8; 15.4\%) using a \textbf{User Performance \textcolor[HTML]{6AB187}{[UP]}} style evaluation, where the performance of users with the system (or independently with the visualizations) was recorded to isolate the effects of specific features. For example, \cite{xiang2019interactive} used logged data as part of an expert case study to compare improvements in label accuracy after iterations of an algorithm. Other studies recorded task completion time and/or accuracy on predetermined tasks given to users to evaluate effectiveness~\cite{zhao2018iforest,ming2018rulematrix}, or scored (via expert ratings, or automatically) the results of an analysis session~\cite{el2018visual,el2019semantic}.
A few papers~\cite{wexler2019if,strobelt2018s} reported usage statistics such as page views from open-source releases.

Ten papers (19.2\%) described \textbf{Algorithm Performance \textcolor[HTML]{6AB187}{[AP]}} evaluations, consisting of quantitative studies of the performance or quality of visualization algorithms or algorithms on which the visualized information depends. A few focused on improving ML model understanding among novices~\cite{kahng2018gan,wang2020cnn}, using an \textit{Evaluating Communication Through Visualization (CTV)} approach focused on how successfully novice end users grasped important concepts as communicated by the tool.

%% file: 06_discussion.tex
\section{Discussion}
Our findings characterize the current scope of VIS4ML research within the broader requirements of human-in-the-loop in ML production. Many of the papers we reviewed make bespoke visualization contributions demonstrated by specific use cases and evaluate them by taking study participants' insights taken at face value. This is not necessarily problematic if the goals of VIS4ML research are exploratory, driven by the purpose of identifying new engineering solutions for surfacing certain information in ML pipelines and reporting lessons learned in the process. This kind of design study methodology has become an established mode for visualization research~\cite{meyer2019criteria}. However, we observed that papers often used their case studies to draw general conclusions about the utility of certain representations or interaction designs for improving the performance of the model or pipeline, such as by making claims about how a system demonstrates the value of integrating human knowledge. This suggests an overlooked distinction between the nature of a design study versus a controlled empirical comparison~\cite{meyer2019criteria}, and between engineering artifacts and scientific knowledge~\cite{hullman2022worst}.  
Here we discuss specific threats to the generalizability of VIS4ML research grounded in our analysis and propose near-term and future-looking recommendations for improving the alignment between claims made in VIS4ML research and the procedures used to validate them.

\subsection{Threats to Generalizability}

\textbf{Ambiguous characterization of human expertise:}
Similar to prior work by Sperrle et al.~\cite{sperrle2021survey}, we also observe strong yet often implicit knowledge assumptions of users of VIS4ML systems. At times, papers' claims about the accessibility of a tool seemed to contradict the knowledge needed to use the system confidently; e.g., one paper presented a hypothetical scenario involving a domain expert who was ``not conversant with complex modeling techniques,'' using a tool that required interpreting residuals and parameter weights. Describing required expertise and reporting sources of confusion and the amount of training needed for experts to use the system successfully (as a few papers did~\cite{alsallakh2014visual,liu2016towards,liu2017visual}) is vital for transparency around dependencies, and will benefit practitioners who try to adopt the work.
\newline
\vspace{-2mm}

\noindent \textbf{Overfitting to specific use cases:}
The predominance of hypothetical usage scenarios and case studies suggests that success is often interpreted as a paper's ability to demonstrate a few (usually 1-3) instances in which a VIS4ML system appears to provide value. A risk of this form of validation is that researchers will inadvertently build tools to confirm their a priori knowledge about some dataset or modeling pipeline but which are overfit to those specific insights. We observed papers often using common datasets that have been thoroughly explored in the ML community, and validating systems by showing that previously identified patterns were also visible using the new system. The problem is that being able to present at least one usage scenario in which a tool is perceived as helpful is different from showing that, on average, using that tool improves some aspect of the ML pipeline. The former takes the form of an existence proof, whereas the latter requires studying a greater number of cases relative to a baseline representing what experts would do without a tool. 
\newline
\vspace{-2mm}

\noindent \textbf{Constrained evaluation practices:}
Overfitting can also arise from heavy reliance on a few experts as consultants in the design process, especially when those same experts are consulted to evaluate a system, as we observed in multiple papers. 
Based on our analysis, papers consistently rely on certain forms of evaluative evidence---namely, a few usage scenarios or case studies with small numbers of experts in which the experts make what they perceive as discoveries or confirm their prior expectations---to validate the work. 
While these practices are not necessarily unreasonable for a burgeoning research area in which researchers invest considerable effort in making tools to support ML practice they may not be intimately familiar with, failure to acknowledge such ``validation gaps'' threatens the transparency and generalizability of VIS4ML research. 
\newline
\vspace{-2mm}

\noindent \textbf{Postulating broader utility from exploratory evidence:}
Helping users draw causal inferences to improve a learning pipeline is a common implicit goal in VIS4ML research. 
However, post-hoc visualizations, even when interactive, are limited in their ability to validate many types of causal hypotheses~\cite{murdoch2019definitions}.
Sometimes papers acknowledged these limitations, such as by calling observations ``speculative''~\cite{rathore2021topoact}, noting that it is difficult to provide specific reasoning about what led to a model decision~\cite{strobelt2018s}, and noting that whether an action could effectively address a perceived problem requires further investigations~\cite{wang2021visual} or statistical evidence~\cite{jaunet2020drlviz}. However, across the papers in our sample such statements were sparse.

Only about one-third of papers reported performance statistics for a usage scenario as evidence that a model had been improved through human involvement. Others relied on experts' statements that a tool was useful through user experience questionnaires or interviews or simply the authors' ability to construct a hypothetical use scenario. While users may be able to confirm propositions about the specific model or pipeline at hand with a VIS4ML tool (e.g., whether removing subsets of data or changing certain hyperparameters affects model output for a certain validation set), hypotheses about the \textit{general} behaviors of an approach cannot be verified without testing under new conditions. These include, for example, hypotheses that a given class of models learns in a particular way or is subject to certain blindspots. A few papers acknowledged the potential for overfitting when users' explanations couldn't be tested on new data~\cite{alsallakh2014visual,das2019beames}. One paper even reported how some expert users of a tool disapproved of the idea of using post-hoc rule extraction to interpret a model's decisions based on potential overfitting and described conditions to avoid this, such as the collection of new data~\cite{alsallakh2014visual}. However, such acknowledgments were rare.

If the goal is to investigate whether human knowledge can \textit{ever} be helpful and how it might be integrated, then papers' motivations should reflect this, for example, by posing questions rather than assertions. However, many papers motivate the work by referring to the demonstrated power of VIS4ML to overcome the shortcomings of fully automated ML pipelines. For example, papers cite prior VIS4ML work in describing how ``As shown by many recent works\ldots, interpreting DNNs with visual analytics has achieved great success''~\cite{wang2021visual}, or ``These visualizations have achieved great success in understanding and analyzing those deep learning models'', referring to visualizations to facilitate developing CNNs, RNNs, GANs, and DQNs~\cite{jin2022gnnlens}. While not necessarily false, such ``success'' is narrowly defined.
\newline
\vspace{-2mm}

\noindent \textbf{Underspecified effort to insights:}
How much difficulty researchers faced in identifying "successful" use cases for a tool is likely to be a useful signal of how reproducibly a tool can improve ML practice. However, papers generally did not describe the selection of specific forms of insights they reported in VDAR-style evaluations, leaving it unclear how unique the scenarios they demonstrated might be. An exception to this is a paper that explicitly noted how the authors explored a large space of possible VDAR processes that the system afforded before deciding on the selected scenarios~\cite{strobelt2017lstmvis}. %: ``we trained and explored many different RNN models, datasets and tasks, including word and character language models, neural machine translation systems, auto-encoders, summarization systems, and classifiers. We also experimented with other types of real and synthetic input data''~\cite{strobelt2017lstmvis}. 
However, it remains unclear whether such exploration was needed to identify successful examples.

\subsection{Recommendations for Action}
Below we summarize near-term and forward-looking aims to address the generalizability gaps listed above. Our recommendations are informed by proposed solutions to larger problems of a lack of rigor and robustness in social science research and may be applicable to the broader area of visual analytics application design. However, our suggestions should be taken as tentative, as it is beyond the scope of our paper to rigorously evaluate these reform proposals as should be expected if reforms are to be effective~\cite{devezer2020case}. 

\subsubsection{Documenting Constraints on Generalizability}
Some of the gaps we highlight can be addressed simply by improving documentation practices about known dependencies in VIS4ML research. Conventional reporting styles in VIS4ML research do not adequately describe the constraints on generality: What dependencies the success of the system may have on the specific configuration of components, users, and other specifics of the setting that was studied. Our results suggest extending prior reporting guidelines for VIS4ML systems~\cite{sperrle2021survey, liu2016towards} to encourage reporting of hidden dependencies, including (1) all datasets used in examples and evaluations, (2) all scalability and memory-related constraints, (3) expectations about how much time users will need to learn the system (and reporting of learning time for expert case studies), and (4) observed sources of confusion and failures among users.
In addition, to help ``close the loop'' researchers should (5) provide model, pipeline, and deployment summary stats before and after changes inspired by using the system. For reproducibility, they should (6) describe parameters and values applied in examples for any parametrized pre-processing steps as well as how those values were reached. Finally, given that VIS4ML claims are demonstrated through specific examples and insights, it is important that the researcher (7) report the nature of their own exploration and insight generation (i.e., time and effort in identifying example insights). Future work should investigate documentation standards and reporting guidelines by taking inspiration from other disciplines such as the social sciences~\cite{simons2017constraints}. 

\subsubsection{Tightening Logical Derivation Chains}
VIS4ML papers should also carefully consider and communicate the deductive logic behind the choices that are made, from motivating the research to defining the conditions of the study. Papers should avoid \textit{loose derivation chains}, to borrow a term applied by Paul Meehl to underspecification in experimental research~\cite{meehl1990summaries}: a lack of evidence of rigorous deductions in moving from theoretical premises to predictions or choices made regarding observed relations. The exploratory nature of VIS4ML research may mean that achieving tight derivation chains is not realistic for many projects. However, papers could be improved by striving to document where choices were made more arbitrarily (e.g., out of convenience) in deciding how to instantiate hypotheses or aspirations in systems. Along these lines, future work might take inspiration from the design study literature, such as in the design activity framework~\cite{mckenna2014design}, to develop similar methods specific to VIS4ML decisions for helping designers reason about and identify best practices in connecting design goals, methods, and outcomes. Future research could extend visualization design and evaluation frameworks (e.g., the nested model~\cite{munzner2009nested}) to emphasize human expertise and prior knowledge captured in VIS4ML design hypotheses and representativeness of said expertise in real-world practice. 

\subsubsection{Bridging from Design Studies to VIS4ML in Practice}
Despite the strength of claims that we observed in the VIS4ML literature, the nature of many of the studies we analyzed appears closer to a design study pattern, in which ``researchers analyze a specific real-world problem faced by domain experts, design a visualization system that supports solving this problem, validate the design, and reflect about lessons learned in order to refine visualization design guidelines''~\cite{sedlmair2012design}. 
Notions of rigor and validation look different in such studies~\cite{meyer2015nested,meyer2019criteria,munzner2009nested}, and expecting replicability and generalizability of VIS4ML research may be premature. While Meyer and Dykes suggest that design researchers ``aim to produce explicit and appropriately scoped expressions of knowledge claims,'' our analysis suggests that properly scoping expressions of knowledge claims may require more concrete guidance to achieve.

In the longer term, VIS4ML research could benefit from forging stronger partnerships with adjacent ML and HCI communities. %Considering the growing interest in using visualizations to scaffold HITL, established types of visualization contributions are likely inadequate. 
As VIS4ML brings visualization research towards the center of data science, ML, and human-centered AI, VIS4ML research should look beyond `insightism'--the superficial reliance on apparent insights produced by use--into pragmatism (usefulness) and cognitivism (impacts on individual and social cognition) to really put the human in the loop~\cite{chen2020isms}. Bridging research to ML practice requires exploring ways to negotiate responsibilities, building deeper research collaborations between visualization and ML researchers (e.g., CARE-ful partnerships~\cite{akbaba2023troubling}), defining boundary objects for knowledge transfer, and addressing the cost and effort of replicating findings in VIS4ML research. 

% \jessica{could this last part be 'addressing the cost and effort of replicating findings in VIS4ML research'? I don't really understand how the cost and effort relates, unless you mean something like how much effort will it take some practitioner to get the same effects in their setting}

%% file: 07_conclusion.tex
\section{Conclusion}
We contribute a focused analysis of 52 VIS4ML papers representing design hypotheses about how integrating human knowledge can help ``close the loop'' in ML practice. We observe a general optimism about the potential for human integration to transform ML practice and research and heavy reliance on collaborations with experts who such tools might help. However, these aspirations are not always accompanied by evaluations demonstrating success in these goals. Our findings show gaps in the generalizability of VIS4ML research contributions, indicating that we are only closing a narrow instantiation of the human-in-the-loop model. We make recommendations for action that include transparent reporting, tightening logical derivation chains in VIS4ML research practices, and extending current design study approaches to ML practices by exploring partnerships with the broader human-centered AI research community.

\section{Acknowledgments}
\label{acknowledement}
We thank our reviewers for their helpful feedback. Hullman is supported by the NSF (IIS-2211939). 